%% file: 00_main.tex
\documentclass[manuscript]{acmart}
\usepackage{xtab}
\usepackage{comment}
\usepackage{array}
\usepackage{graphicx}
\usepackage{subcaption}
\usepackage{csquotes}

\AtBeginDocument{%
  \providecommand\BibTeX{{%
    \normalfont B\kern-0.5em{\scshape i\kern-0.25em b}\kern-0.8em\TeX}}}

\renewcommand\footnotetextcopyrightpermission[1]{} 
\acmConference[]{Submitted to MUM 2024}{Dec 1–-4, 2024}{Stockholm, Sweden}

\begin{document}

\title[Trickery: Exploring a Serious Game Approach to Raise Awareness of Deceptive Patterns]{Trickery: Exploring a Serious Game Approach to Raise Awareness of Deceptive Patterns}

\author{Kirill Kronhardt}
\email{kirill.kronhardt@tu-dortmund.de}
\affiliation{%
  \institution{TU Dortmund University}
  \department{Inclusive Human-Robot Interaction}
  \orcid{}
  \streetaddress{Emil-Figge-Straße 50}
  \city{Dortmund}
  \state{NRW}
  \country{Germany}
}

\author{Kevin Rolfes}
\email{k.r.whs@outlook.de}
\affiliation{%
  \institution{Westphalian University of Applied Sciences}
    \orcid{}
  \streetaddress{Neidenburger Straße 43}
  \city{Gelsenkirchen}
  \state{NRW}
  \country{Germany}
}

\author{Jens Gerken}
\email{jens.gerken@tu-dortmund.de}
\affiliation{%
  \institution{TU Dortmund University}
  \department{Inclusive Human-Robot Interaction}
  \orcid{}
  \streetaddress{Emil-Figge-Straße 50}
  \city{Dortmund}
  \state{NRW}
  \country{Germany}
}

\renewcommand{\shortauthors}{Kronhardt et al.}

\begin{abstract}
Deceptive patterns are often used in interface design to manipulate users into taking actions they would not otherwise take, such as consenting to excessive data collection. We present \emph{Trickery}, a narrative serious game that incorporates seven gamified deceptive patterns. We designed the game as a potential mechanism for raising awareness of, and increasing resistance to, deceptive patterns through direct consequences of player actions. We conducted an explorative gameplay study to examine player behavior when confronted with the game \emph{Trickery}. In addition, we conducted an online survey to shed light on the perceived helpfulness of our gamified deceptive patterns.
 Our results reveal different player motivations and driving forces that players used to justify their behavior when confronted with deceptive patterns in the Trickery game. In addition, we identified several influencing factors that need to be considered when adapting deceptive patterns into gameplay. Overall, the approach appears to be a promising solution for increasing user understanding and awareness of deceptive patterns.
\end{abstract}


\received[Submitted to MUM 2024 on]{August 22nd, 2024}

\maketitle
\input{01_introduction}

\input{02_related-work}
\input{03_game-adapted-dark-patterns}
\input{04_exploratory-study}
\input{04_survey-study}

\input{05_discussion}
\input{06_conclusion}

\begin{acks}
  \subsection{Funding}
  The research is supported by the Federal Ministry of Education and Research for Education and Research of the Federal Republic of Germany under the funding codes FKZ: \href{https://foerderportal.bund.de/foekat/jsp/SucheAction.do?actionMode=view&fkz=16KIS1631}{16KIS1631}, FKZ: \href{https://foerderportal.bund.de/foekat/jsp/SucheAction.do?actionMode=view&fkz=16KIS1629}{16KIS1629}. The responsibility for the for the content of this publication lies with the authors.

  \subsection{Use of Generative AI}
  Generative AI was used to improve the quality of existing text written by the authors. The authors acknowledge that the resulting Work in its totality is an accurate representation of the authors’ underlying work and novel intellectual contributions and is not primarily the result of the tool’s generative capabilities. The authors accept responsibility for the veracity and correctness of all material in their Work, including any computer-generated material according to the ACM policy on authorship\footnote{ACM Policy on Authorship: \url{https://www.acm.org/publications/policies/frequently-asked-questions}}.
\end{acks}

\bibliographystyle{ACM-Reference-Format}
  \bibliography{PETS24-player-beware}

\end{document}

%% file: 01_introduction.tex
\section{Introduction}

\subsection{Introduction}
    In the current digital landscape, website providers play a dual role: they offer free and important services to users, such as social media platforms, while quietly capitalizing on the data they collect about their users through targeted advertisements. This excessive data collection can lead to tangible harm to users through invasion of privacy, financial loss, and cognitive burden, as well as harm to the collective welfare through unfair competition or unanticipated societal consequences~\cite{Mathur.05062021}.
    While users express concern about their information privacy in interviews and questionnaires, users do little to actually protect their data when observed~\cite{acquisti2005privacy, acquisti2015privacy, van2015older}. Users are often unaware of the extent of the information collected about them~\cite{farke2021privacy, vervier2017perceptions}. Despite regulatory barriers like the EU General Data Protection Regulation (GDPR), which requires explicit informed consent from users to collect data, many users give consent through cookie consent banners quickly to get to the websites' content, disregarding potential privacy risks~\cite{anderson2020gdpr, giese2022factors}.
    
    An important factor in influencing users' decision to consent to data collection is interface design~\cite{Utz.11062019}. Specifically, website providers employ manipulative interface design strategies, so-called deceptive patterns~\footnote{In prior research, deceptive patterns have mostly been called dark patterns. However, we acknowledge that the ACM Diversity and Inclusion Council now includes the term “dark patterns” on a list of problematic terms (https://www.acm.org/diversity-inclusion/words-matter). As a consequence, multiple leading authors in the field have started to adopt the less ambigious term deceptive patterns.}, to increase the likelihood that users give consent to data collection \cite{Gray.26.04.2018, BongardBlanchy.06282021} or extend their stay on a website. Therefore, deceptive patterns trick users into performing unintended and unwanted actions by exploiting psychological biases, such as prompting \emph{System 1 thinking} (fast, unconscious, automatic, less rational), and humans' fundamental need to belong \cite{Bosch.2016}. Similarly to the lack of awareness about the amount of collected data, users are either unaware of deceptive patterns \cite{DiGeronimo.04212020} or are aware that they are being manipulated but do not understand how deceptive patterns can concretely harm them \cite{BongardBlanchy.06282021}. To counteract deceptive patterns \citeauthor{BongardBlanchy.06282021} propose raising awareness, facilitating detection, and bolstering resistance towards deceptive patterns through educational, design-oriented, technological or regulatory measures \cite{BongardBlanchy.06282021}. Similarly, \citeauthor{lu2023awareness} propose that targeting user awareness and user action can lead to end-user-empowerment, bolstering resistance against deceptive patterns \cite{lu2023awareness}. 
    Our paper takes an educational approach to counteracting deceptive patterns, by exploring serious games to raise awareness of deceptive patterns.  
    Serious games are games that are used for non-entertainment purposes \cite{kara2021systematic}. There have been a number of serious games aimed at educating people about privacy and cyber security awareness or even deceptive patterns \cite{Berger.2019, Hart.2020, Maragkoudaki.2022, Akinyemi.2022}. Games inherently increase motivation and fun and thereby improve learning, i.e. induce a change in behavior, attitude, health, understanding, or knowledge \cite{becker2021s}. In their meta-analysis of serious games used in education \citeauthor{zhonggen2019meta} found that serious games encouraged longer engagement with the topic at hand and learners were generally more motivated than in nongame-based learning approaches~\cite{zhonggen2019meta}. Serious games in the context of privacy awareness can take on a variety of forms, often aiming to train users directly to detect or understand certain privacy issues. For example, the \emph{Cookie consent speed run} game combines realistic cookie consent forms with game mechanics such as rewards and competition\footnote{e.g., https://cookieconsentspeed.run/}. However, we know from educational research, that such direct approaches, which do not require much reflection and thought on the users' part, may mostly trigger short-term memory learning and not allow the creation of new neural links for reflection and understanding \cite{Liston_1994,alterio2003learning}. We were therefore looking for approaches which could trigger a deeper level of reflection beyond rote memorization of deceptive patterns with the aim to create overall awareness and resistance.

    In this paper, we present \emph{Trickery}, a serious game which is based on narrative storytelling to raise awareness about deceptive patterns. The gameplay is designed around seven gamified deceptive patterns, i.e., gameplay adaptations of real deceptive patterns as found in the literature~\cite{Gray.26.04.2018}. We then conducted an explorative laboratory based gameplay study (Study \#1) and a subsequential Online Survey Study (Study \#2) with the following guiding research questions:

    \begin{enumerate}
             \item \textbf{RQ1:} What are \emph{Influencing Factors}, which need to be considered for the design of such gamified deceptive patterns so that they can be effective to raise awareness? (Study \#1)
        
        \item\textbf{RQ2:}What are the \emph{Motivations and Driving Forces} which might help to explain the behavior of users playing the game and falling or resisting a certain gamified deceptive pattern? (Study \#1)

        \item \textbf{RQ3:} Do users actually find our gamified deceptive patterns helpful to understand the real deceptive pattern concepts from the literature? (Study \#2)
    \end{enumerate}

%% file: 02_related-work.tex
\section{Related Work}
 \subsection{Taxonomies of Deceptive Patterns}
As far back as 2010, Conti and Sobiesk described malicious interface designs and called for joint work between the security and human-computer interaction communities to address this issue \cite{Conti.2010}. In addition, their paper, based on three extensive surveys, formalized a first taxonomy of different types of deceptive patterns, which served as the foundation for further research in this area. Another early example of a taxonomy of deceptive patterns was presented by \citeauthor{brignull2015dark}, who was among the first to raise awareness not just in the research community, but among designers and users through his website on deceptive patterns \cite{brignull2015dark, Brignull.13.09.2023}. Quickly, further research emerged that aimed at gaining a deeper understanding of what deceptive patterns are and how to properly classify them.  In 2013, \citeauthor{Zagal1043332} first applied the concept of deceptive patterns to the design of games, coining the term "Dark Game Design Pattern" and defining it as "a pattern used intentionally by a game creator to cause negative experiences for players which are against their best interest and likely to happen without their consent" \cite{Zagal1043332}. Their exemplary deceptive game design patterns include temporal deceptive patterns, such as \emph{Grinding}, monetary deceptive patterns, such as \emph{Pay-to-Skip}, and social capital-based deceptive patterns, such as  \emph{Social Pyramid Schemes}. These kinds of deceptive patterns, as well as psychological deceptive patterns, are even found in mobile games for children from 0-5 years old~\cite{sousaDarkSideFun2023}. More recently, \citeauthor{aagaardGameDarkPatterns2022} performed interviews and workshops with both players and designers of mobile games, finding that developers are often preassured into integrating deceptive patterns into their games to drive engagement~\cite{aagaardGameDarkPatterns2022}. \citeauthor{Bosch.2016} directly aligned their classification to Hopeman's privacy design strategies \cite{Bosch.2016, Hoepman.2014}. The idea was that privacy deceptive strategies, the underlying basis for privacy deceptive patterns, could be described as a direct reversal of privacy strategies, originally aimed at increasing data privacy. The result is a rather abstract set of terms, such as \emph{maximize} or \emph{obscure}. Building on this, in what is regarded as one of the most influential deceptive pattern taxonomies to this day, \citeauthor{Gray.26.04.2018} were able to provide a balance of abstraction and over-specification  \cite{Gray.26.04.2018}. Their taxonomy consists of five main categories, which manage to keep the same level of abstraction and at the same time be concrete enough that the terms used allow readers to relate them to specific examples. 
A slightly different approach was presented by \citeauthor{Mathur.2019}, who classified different deceptive patterns with respect to their influence on user decision-making \cite{Mathur.2019}. The authors describe their approach as "offering a set of shared higher-level attributes that could descriptively organize instances of deceptive patterns in the literature" \cite{Mathur.05062021}. Most recently and \textit{after our gameplay design and studies were already concluded,} \citeauthor{gray_ontology_2024} presented an ontology about deceptive patterns with the aim to unify the different perspectives and provide a clear understanding of the relationship between the varying perspectives~\cite{gray_ontology_2024}.

\subsection{Understanding Users' Vulnerability to Deceptive Patterns}
Aiming to understand why users actually are vulnerable to deceptive patterns, different theories and conceptual models have been applied. \citeauthor{Xiao2011} identified affective and cognitive mechanisms with certain deceptive information practices and created an overall theoretical model \cite{Xiao2011}. \citeauthor{Bosch.2016} applied Kahnemann's Dual process theory of System 1 and System 2 thinking \cite{Kahneman.2011}, which is based on the understanding that humans have two modes of thinking, a fast one (System 1, unconscious, automatic, less rational) and a slow one (System 2, conscious, rational). The assumption of \citeauthor{Bosch.2016} is that deceptive patterns systematically exploit users' System 1 thinking. \citeauthor{Lewis.2014} tried to establish a link between deceptive patterns and psychological motivators, based on Reiss's Desires theory \cite{reiss2004multifaceted}. 

Several studies explored user perceptions when faced with deceptive patterns. \citeauthor{Luguri.2021} found that mild deceptive patterns are more effective and less educated users are more susceptible to deceptive patterns~\cite{Luguri.2021}. In subscription decisions, decision architecture appeared to be even more important than material differences like price~\cite{Luguri.2021}. While moderately aware of deceptive patterns, users have a resigned attitude towards them, as they believe themselves dependent on the very services that employ deceptive patterns~\cite{maierDarkDesignPatterns2020}. Users may struggle to identify deceptive patterns~\cite{DiGeronimo.04212020} but even if users are aware of and able to identify deceptive patterns, they lack knowledge of specific harms deceptive patterns might cause and how to oppose them~\cite{BongardBlanchy.06282021}. More recently, \citeauthor{mildnerDefendingDarkArts2023} developed and evaluated a five-question tool to help users assess the presence of deceptive patterns within social media interfaces~\cite{mildnerDefendingDarkArts2023}. While the results show promise in that a majority of users were able to discern deceptive interfaces from non-deceptive ones, ratings for the \enquote{darkness} of the deceptive interfaces remained considerably low, indicating some difficulty~\cite{mildnerDefendingDarkArts2023}. 

\subsection{Serious Games to Raise Awareness for Data Privacy Issues}
\label{sec:relatedwork_games}
\citeauthor{becker2021s} defines serious games as "games designed specifically for purposes other than or in addition to pure entertainment" \cite{becker2021s}. \citeauthor{alvarez2011introduction} further specify that such games may include aspects of tutoring, teaching, training, communication, or information \cite{alvarez2011introduction}. Games, as such, can in essence be defined as consisting of a closed environment that is interactive, has a set of specific rules, and has one or more goals for the player \cite{becker2021s}. Serious games have been shown to increase learning outcomes for visual and spatial processing, complex concepts and abstract thinking, and deduction and hypothesis testing \cite{dondlinger2007Educational}. However, serious games can not only increase learning outcomes. In the form of persuasive games, they can be strongly effective at promoting behavior change \cite{ndule2023games}. In our recent work, we argue that serious and persuasive games are a promising solution to bolster resistance against deceptive patterns, with the specifics of game design, game presentation, and study design being crutial aspects to consider for an effective outcome~\cite{kronhardtStartPlayingSerious2024}.

The idea of serious games being effective for learning can be ascribed to multiple aspects. For one, we know that the experience of situations with direct consequences based on decisions should result in emotional learning \cite{schiebener2015decision}. Serious games provide such an environment, where, through game design, decisions can lead to immediate consequences. In the context of data privacy, we also know of the importance of immediate consequences. The less prominent risks are, the more likely users will disclose information \cite{krasnova2012self}. A game also provides the possibility to include failures and offer a learning experience from failure. Again, the literature shows that such experiences, associated to negative emotions, can be highly influential \cite{rogers1975protection}, and accordingly, negative emotions linked to privacy violations may also trigger privacy protection behavior \cite{schwael2018}. This persuasive strategy of allowing players to observe the cause and effect of their behaviors is called \emph{Simulation}, and is one of the most used persuasive strategies in persuasive games \cite{naul2020story}.

On the other hand, narrative elements increase immersion in and engagement with a serious game, as well as motivate players to learn more compared to serious games without narrative elements \cite{naul2020story}. \citeauthor{naul2020story} also find that intrinsic integration of narrative, game-play and learning content provide both educational and motivational benefits to serious games~\cite{naul2020story}. Such motivational benefits of intrinsic integration were also identified by \citeauthor{habgoodMotivatingChildrenLearn2011a}~\cite{habgoodMotivatingChildrenLearn2011a}. Moreover, they identify that humans interact with virtual agents much the same way as they do with real people, including developing emotional responses (including negative ones) towards these agents \cite{naul2020story}.

Existing serious games in the data privacy context have approached the concept quite differently. \citeauthor{Akinyemi.2022} proposed the game "Dark Cookie" which aims to train users to spot deceptive patterns in cookie banners \cite{Akinyemi.2022}. To do so, it embeds cookie banners in a kind of cover story (four bears and a raccoon) where the game tries to trick the user into accepting deceptive cookies. PrivaCity is a chatbot game with a focus on smart cities \cite{Berger.2019}. It is similar to older text-based adventures and offers the users certain choices related to data privacy in a ficticious scenario of a smart city. 

The approach by \citeauthor{Gupta.2020} which aims to train cyber security professionals falls between the categories of serious games and gamification \cite{Gupta.2020}. Their approach confronts players with realistic threat scenarios of fishing and threat hunting and adds gamification elements such as a scoreboard and time sparsity.  

\citeauthor{Maragkoudaki.2022} explored the idea of a virtual reality escape room to provide an environment for privacy awareness \cite{Maragkoudaki.2022}. While the overall escape room story is unrelated to privacy at first sight, the authors slipped in specific data privacy examples, such as very long privacy policies or a computer search history page. Depending on the players' behavior, they get rewarded by receiving more time to leave the escape room and solve the riddles. In addition, it confronts the player with messages in a subtextual form, e.g. written on walls, that aim to raise privacy awareness.

\citeauthor{Hart.2020} follow yet again a very different approach by developing the physical tabletop game "Riskio" which, however, again adapts a very specific data privacy scenario and allows players to explore it in typical boardgame style with game mechanics such as card decks and different turn-based game phases \cite{Hart.2020}. In a similar, physical approach, \citeauthor{tjostheimSeriousGameApps2024} introduces a board game called "Dark Pattern" designed to raise awareness about deceptive data-sharing practices in apps~\cite{tjostheimSeriousGameApps2024}. Players are tasked with installing apps while trying to minimize personal data shared. \citeauthor{tjostheimSeriousGameApps2024} found that while knowledge about deceptive patterns increased, the impact on behavioral intention to protect privacy remained weak~\cite{tjostheimSeriousGameApps2024}. Keeping with the board game aspect, \citeauthor{nyvollSeriousInteractiveBoard2020} designed an interactive social deduction board game in which players try to deduce which player is a \enquote{CEO} who is deploying deceptive patterns while they are using a smartphone.

Overall, all these games have in common that the implemented narrative, scenario, or privacy-related game situations provide a \textit{direct adaptation} of the real-world situation, most often by placing the real-world interface component (i.e., a cookie banner, a deceptive pattern), quite literally, in a game environment. While such learning environments can show positive effects in the short term, we also know from educational research, that approaches which require more reflection and thought and involve powerful emotional experiences might be more effective long term, through the creation of new neural links for reflection and understanding~\cite{Liston_1994,alterio2003learning}. 

In contrast to previous approaches, we aim to design a narrative-driven game with the following attributes, based on the lessons learned in the literature:
\begin{itemize}
    \item a compelling narrative to increase motivation and educational value \cite{naul2020story}
    \item a simulation environment to allow the player to experience the immediate consequences of their actions to trigger emotional learning \cite{schiebener2015decision}.
    \item ways for powerful emotional experiences and responses, such as interaction with a virtual agent, to facilitate reflection and understanding \cite{naul2020story, alterio2003learning}
    \item an intrinsic integration of deceptive patterns into the game world \cite{habgoodMotivatingChildrenLearn2011a}.
\end{itemize}

In addition, we were interested in a gameplay concept which provides utmost flexibility with regards to how deceptive patterns could be adapted within the game. By this we mean that the gameplay concept should lend itself towards the possibility to implement the same deceptive pattern in different forms. For the short term, this allowed us to a) reduce the need for compromise when designing a gamified deceptive pattern to make it fit into the overall narrative and gameplay concept, b) provide the basis for future research and iterations in design, i.e., we may be able to keep the gameplay concept in future version of the game while still redesigning or exchanging the specific adaptations of a deceptive pattern.



%% file: 03_game-adapted-dark-patterns.tex
\section{Trickery Gameplay and Deceptive Pattern Design}
\label{sec:game-adapted-dark-patterns}
For the sake of clarity and to avoid confusion, we will first define the following key terms that are critical to understanding the sections that follow:

\begin{enumerate}
     \item \emph{Deceptive Pattern Concept:} The abstract idea or general definition of a deceptive pattern, discussing its characteristics and how it works, e.g., "Aesthetic manipulation is any manipulation of the user interface that deals more directly with form than function" \cite{Gray.26.04.2018}. 
     \item \emph{Deceptive Pattern Example:} Concrete instances or implementations of deceptive pattern concepts found in real-world scenarios, such as the specific use of "aesthetic manipulation" in a cookie consent banner on a website.
     \item \emph{Gamified Deceptive Pattern:} Adaptation of a deceptive pattern concept for use within \emph{Trickery}. It describes how a deceptive pattern concept can be represented in a game for educational purposes.
     \item \emph{Enriching Game Mechanics:} Game mechanics that are used to make the game more fun and enhance the overall game experience, but that have no direct relationship to a deceptive pattern concept.
\end{enumerate}

To aid in the design and analysis of serious games that educate users about deceptive patterns, we introduce the concept of gamified deceptive patterns. These aim to adapt deceptive pattern concepts for use in a serious game such as Trickery by a) providing a way in which the game manipulates players, and b) showing the negative consequences of falling for the manipulation during gameplay. In addition, they may also need to include a countermeasure, i.e. a way for the user to overcome the manipulation in order to progress in the game. Therefore, ideally, when a player experiences such a gamified deception pattern, he or she will build up negative emotions about being manipulated (increasing awareness of deceptive patterns when they encounter them) and learn to look for possible countermeasures, otherwise he or she will not be able to progress in the game (strengthening resistance).

\subsection{Gameplay Concept}
We were looking for a gameplay approach that would rely on a deceptive mechanism at its core and combine it with strong and emotional storytelling. As already discussed, current approaches in the context of serious games to raise awareness (see \ref{sec:relatedwork_games}) do not integrate these aspects. Therefore, we also researched several commercial games with narrative approaches that might lend themselves to the integration of deceptive patterns. We conducted a small-scale ideation workshop with four participants experienced with video games to identify possible game design approaches and related video games. In particular, we found the games \emph{The Stanley Parable\footnote{https://www.stanleyparable.com/, last accessed \today}}, \emph{Portal\footnote{https://www.thinkwithportals.com/, last accessed \today}}, and \emph{Bioshock\footnote{https://store.steampowered.com/app/7670/BioShock/, last accessed \today}} to provide both deceptive core mechanisms and strong storytelling. In each of these games, players play a silent protagonist who explores their environment from a first-person perspective, completing tasks that eventually lead them to leave the environment. They are accompanied by another character who guides them through the environment but, at least initially, is not visible or otherwise accessible to the player, communicating with them only through audio. This non-playable character (NPC) -- "the narrator" in \emph{The Stanley Parable}, "GLaDOS" in \emph{Portal}, and "Atlas" in \emph{Bioshock}) -- comments on the players' actions and gives them tasks to complete. In each of these games, the companion character has some form of manipulative intent or behavior toward the playable character that only becomes apparent as the game progresses. This allowed us not only to adapt the gameplay for \emph{Trickery}, but also to draw inspiration from the subtle design of such a manipulative character. 

In \emph{Trickery}, the player takes on the role of a new employee in a laboratory who is supposedly being guided through the onboarding process by the lead scientist (hereafter referred to as "the narrator"). The lead scientist is invisible, but communicates with the player via ubiquitous speakers. However, this onboarding process is only a facade. The narrator's true motive is to keep the player character in the lab as long as possible, using deceptive patterns to accomplish this goal. Each room in the game uses a specific gamified deception pattern and can only be completed by finding the appropriate countermeasure. This gameplay concept incorporates a narrative representation of manipulative website vendors and provides a flexible framework in that rooms representing deceptive patterns can be added or removed to teach players about different deceptive pattern concepts or to compare different implementations of the same adaptive pattern concept.

\subsection{Gamified Deceptive Pattern Design}
\label{sec:gamified-patterns}
In this section, we will present our gamified deceptive patterns along the specific implementation in the game \emph{Trickery}. We also aim to provide a more general description to facilitate adaptation to other serious game contexts. In total, we present seven gamified deceptive patterns in order of appearance as implemented in the game \emph{Trickery}. They are based on the deceptive pattern concepts introduced by \citeauthor{Gray.26.04.2018}\cite{Gray.26.04.2018} in 2018. We chose these deceptive pattern concepts because they "serve as strategic motivators for designers" \cite{Gray.26.04.2018}, which fits the narrative of a senior scientist designing a lab that keeps its new hires inside for as long as possible. The concepts of deceptive patterns introduced by \cite{Gray.26.04.2018} have also been widely used as a taxonomy of deceptive patterns by researchers (\cite{voigt2021dark,chromik2019dark,Luguri.2021} to name a few). They are more general than, for example, the privacy dark patterns presented by \cite{Bosch.2016}. \cite{Bosch.2016} or the game design dark patterns of \citeauthor{Zagal1043332} \cite{Zagal1043332}. Thus, they helped us not to limit ourselves to very specific real-world implementations of deceptive patterns, but to aim to convey the underlying concept in our gamified adaptations. \citeauthor{Gray.26.04.2018} introduced five overall deceptive pattern concepts: \emph{Nagging}, \emph{Obstruction}, \emph{Sneaking}, \emph{Interface Interference}, and \emph{Forced Action}, where Interface Interference is an umbrella catergory and consists of three specific deceptive pattern concepts, \emph{Hidden Information}, \emph{Preselection}, and \emph{Aesthetic Manipulation}, which brings us to the total number of seven deceptive pattern concepts. While \citeauthor{Gray.26.04.2018} mapped specific deceptive patterns from an existing taxonomy \cite{brignull2015dark} as well as some more general examples  to their new definitions, we only focused on the deceptive patterns specifically identifiead as strategic motivators in \citeauthor{Gray.26.04.2018}. While the recently published unifying ontology by \citeauthor{gray_ontology_2024} was published a bit too late for our work, it not only builds upon the earlier work from \citeauthor{Gray.26.04.2018} \cite{Gray.26.04.2018} but also clearly links the seven deceptive pattern concepts used in our work to the new ontology~\cite{gray_ontology_2024}. We discuss the specifics of this mapping and how it affects future work regarding \emph{Trickery} in \autoref{sec:limitations}.

\begin{figure}
\centering
\begin{subfigure}{0.45\textwidth}
    \includegraphics[width=\textwidth]{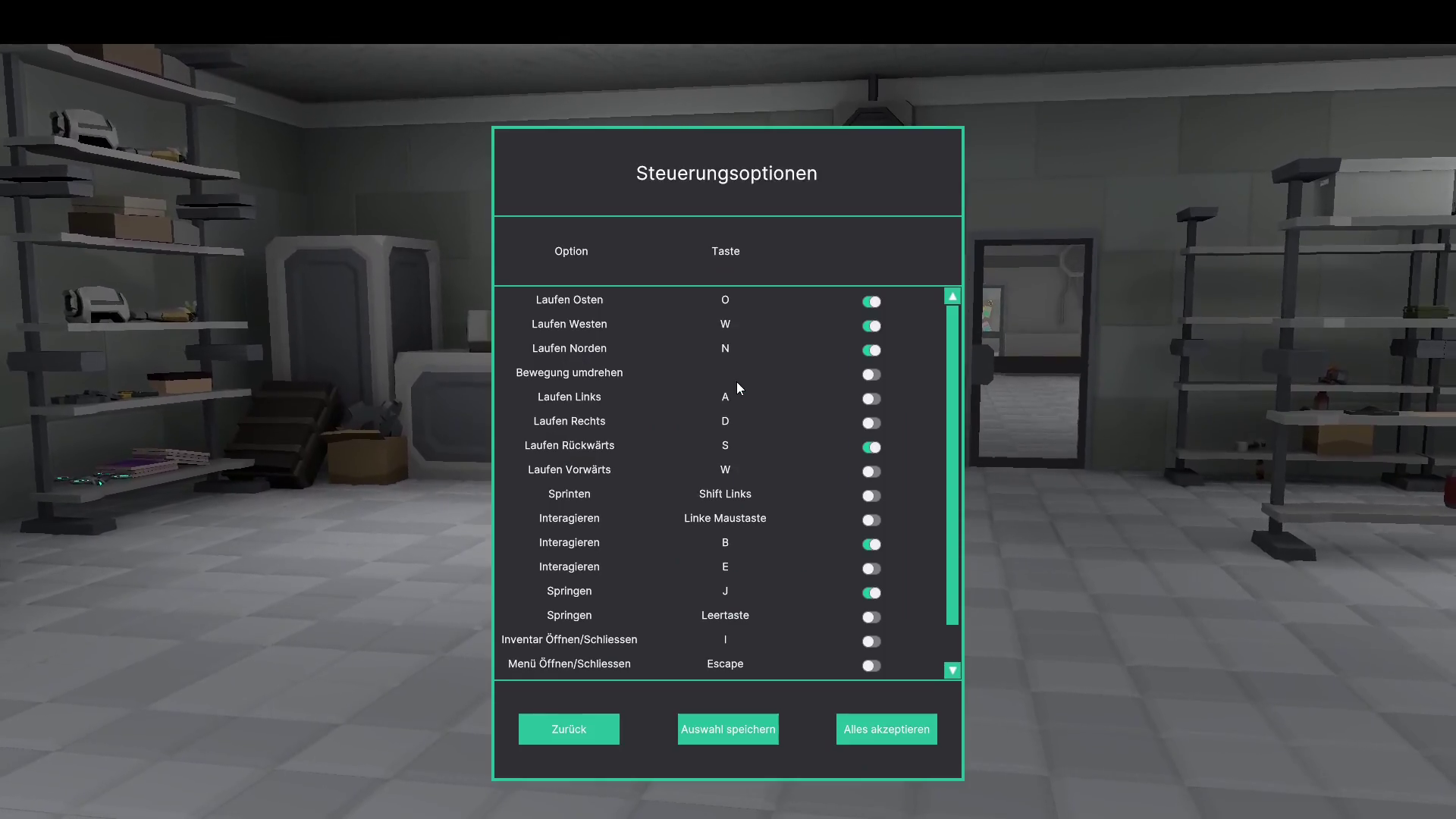}
    \Description[A configuration menu screen to configure control key assignments representing the insensible key mapping deceptive pattern]{A screenshot of the game showing an office storage space. In front is a 2D graphical configuration menu which allows the user to reconfigure the key mapping to control the game. It's representing the insensible key mapping deceptive pattern.} 
    \caption{Preselection: Insensible Key Mapping}
    \label{fig:insensible-key-mapping}
\end{subfigure}
\begin{subfigure}{0.45\textwidth}
    \includegraphics[width=\textwidth]{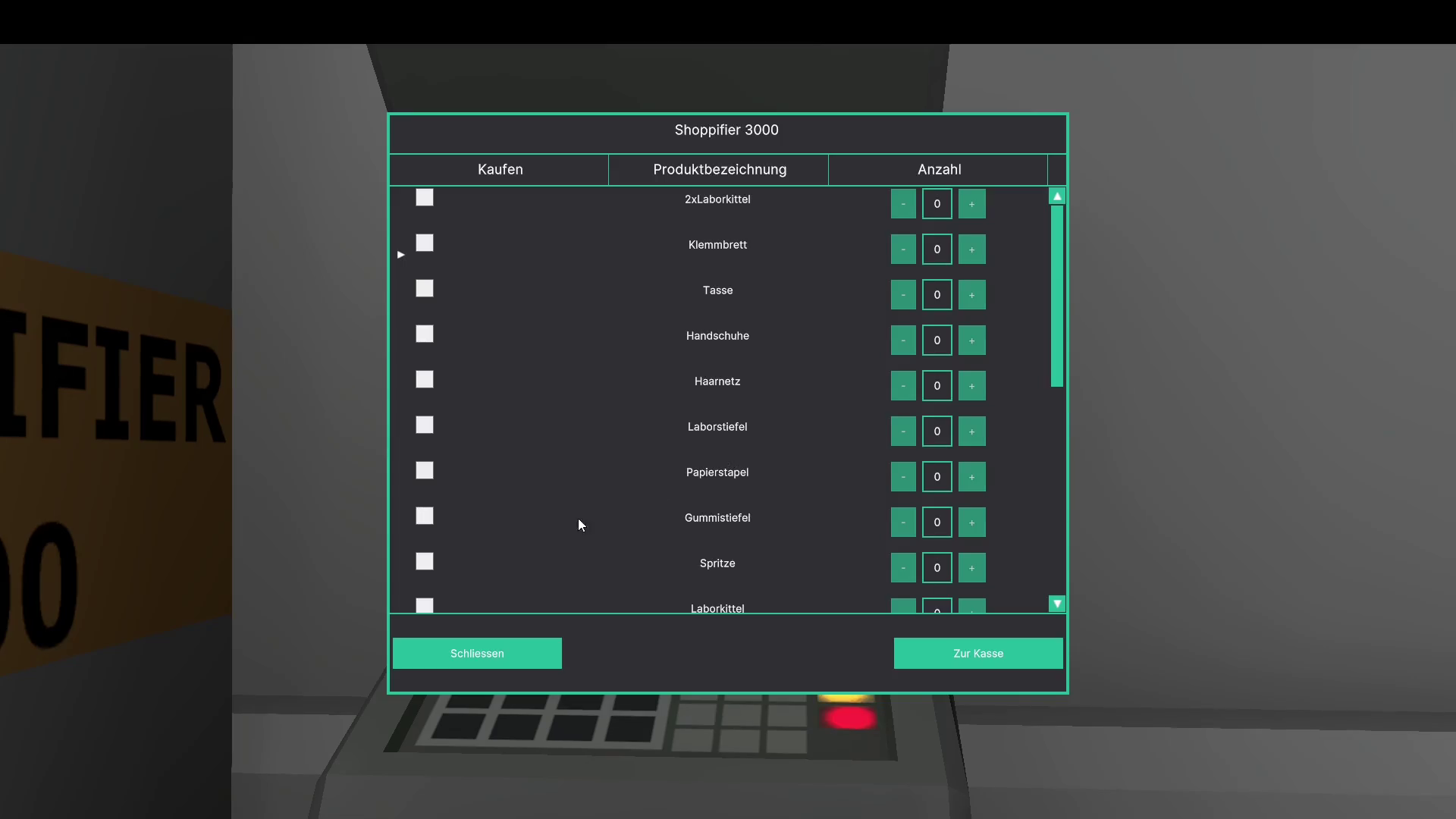}
    \Description[A shopping cart screen with multiple items and controls to select the amount, representing the sneaky shop deceptive pattern]{A screenshot of the game showing a shopping machine in the background. In front is a 2D graphical shopping cart screen, similar to e-commerce webshops, where the user can select which items they want to buy. It's representing the sneaky shop deceptive pattern.}
    \caption{Sneaking: Sneaky Shop}
    \label{fig:sneaky-shop}
\end{subfigure}
\begin{subfigure}{0.45\textwidth}
    \includegraphics[width=\textwidth]{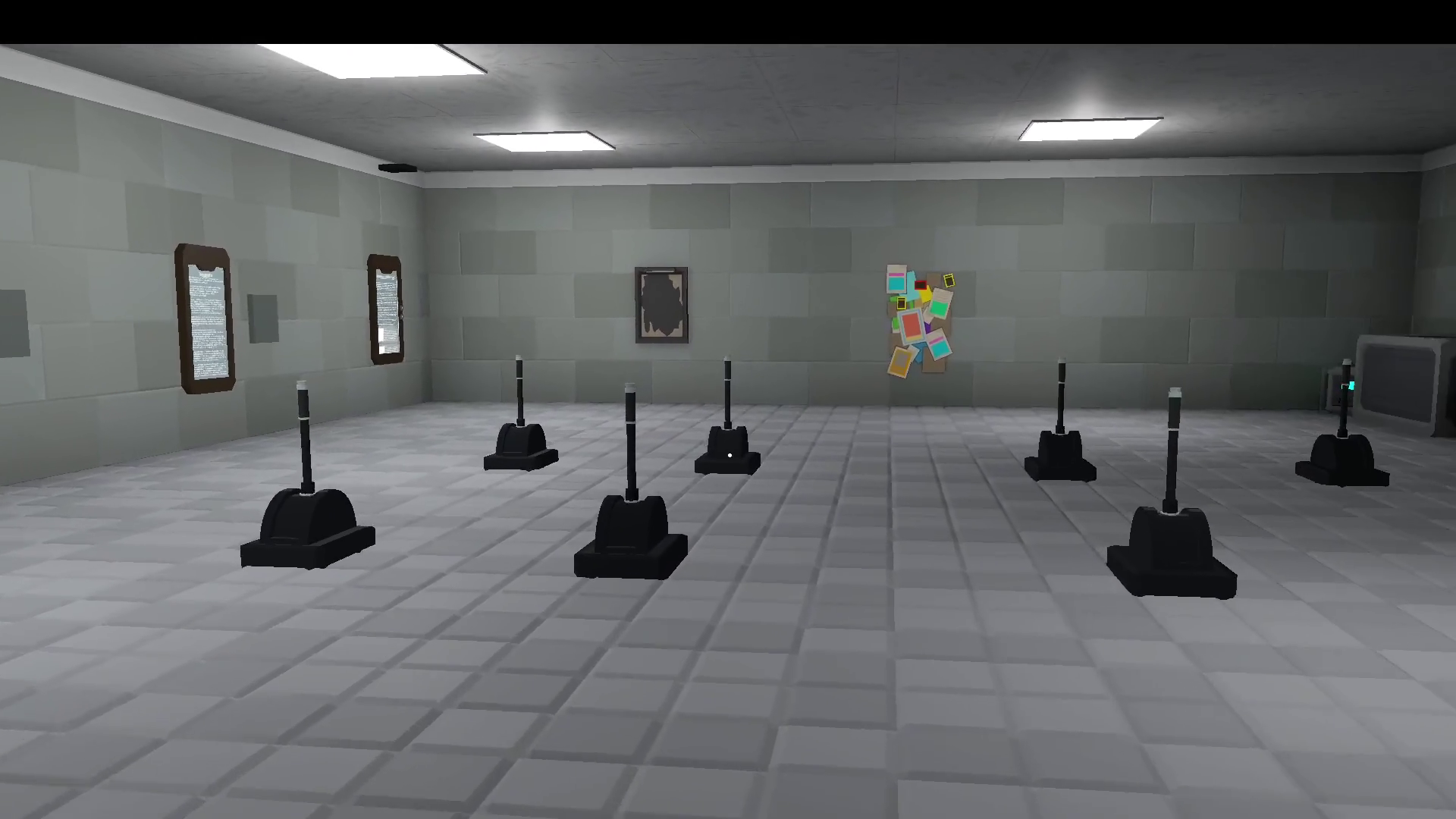}
    \Description[A room with virtual screens hanging on walls which show lots of text]{A screenshot of the game showing a room with multiple screens on the wall and large control levers on the floor. It's representing the walls of text deceptive pattern.}
    \caption{Hidden Information: Walls of Text}
    \label{fig:walls-of-text}
\end{subfigure}
\begin{subfigure}{0.45\textwidth}
    \includegraphics[width=\textwidth]{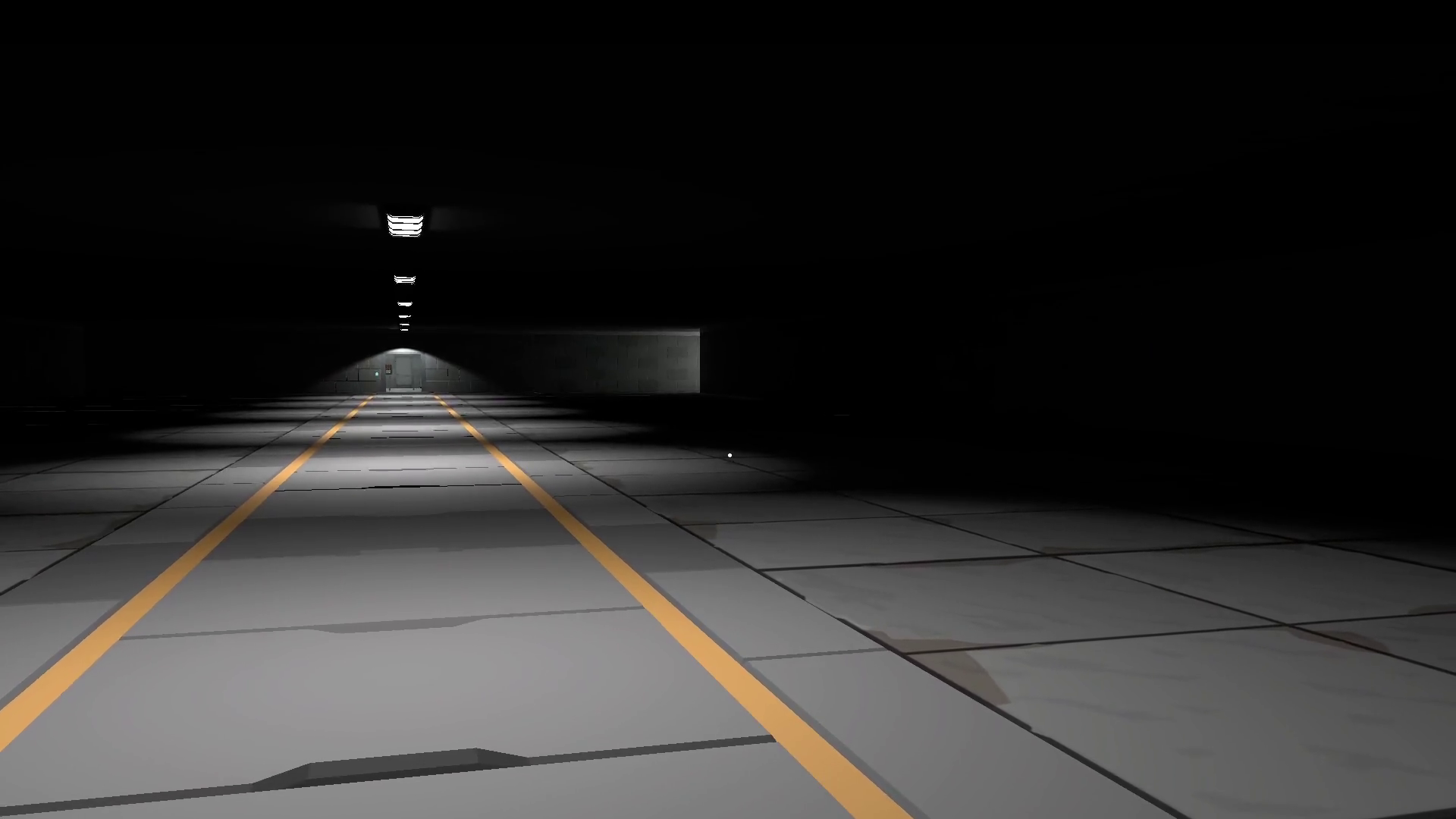}
    \Description[A large room illuminated only in the middle by spotlights]{A screenshot of the game showing a room where only a path in the middle is illuminated, the rest of the room is dark. It's representing the winding hallway \& shortcut deceptive pattern.}
    \caption{Aesthetic Manipulation: Winding Hallway \& Shortcut}
    \label{fig:winding-hallway}
\end{subfigure}
\begin{subfigure}{0.45\textwidth}
    \includegraphics[width=\textwidth]{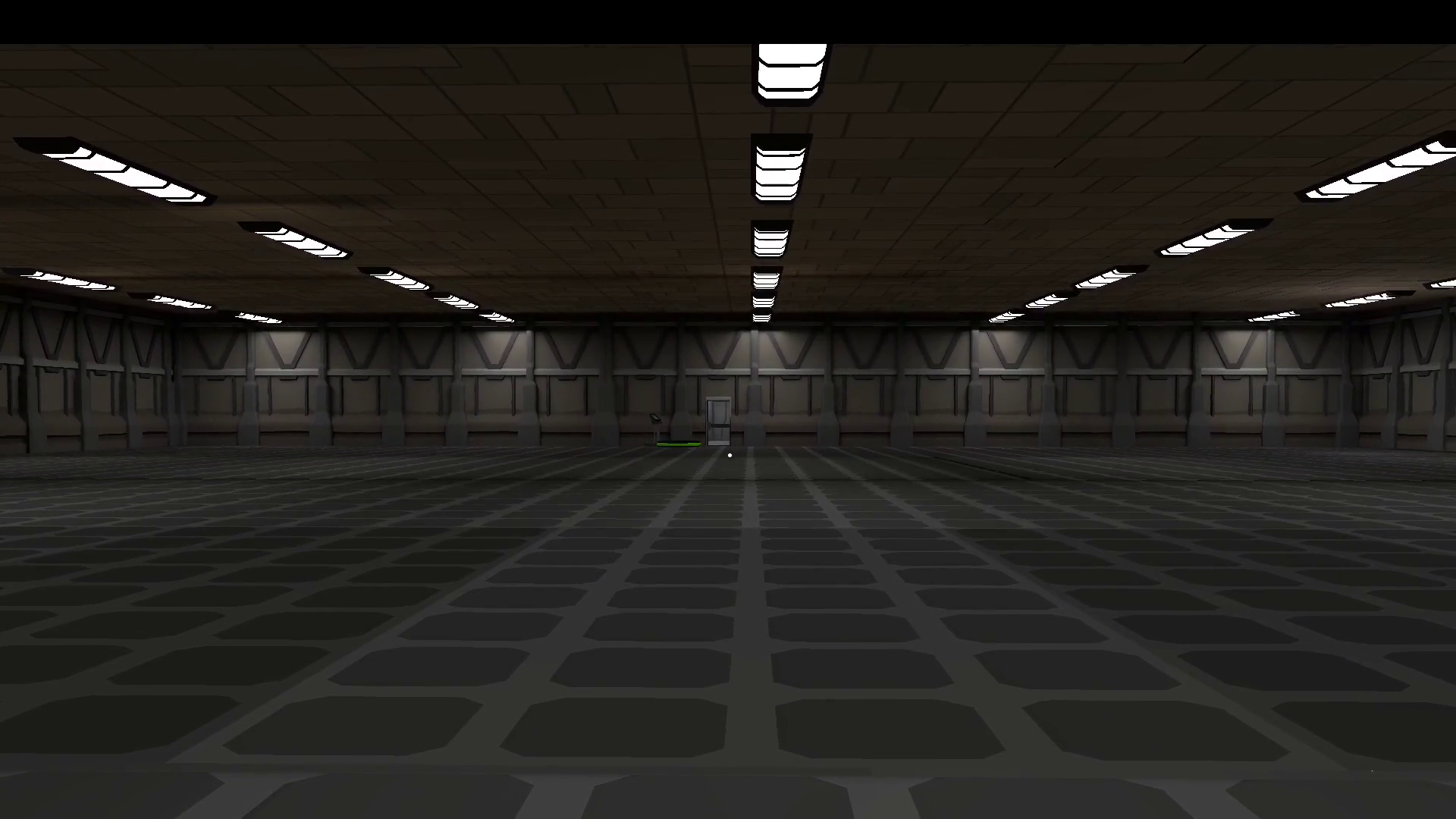}
    \Description[A seemingly empty room which the player has to cross]{A screenshot of the game showing an empty room, slightly illuminated. It's representing the obstacle onslaught deceptive pattern}
    \caption{Obstruction: Obstacle Onslaught}
    \label{fig:obstacle-onslaught}
\end{subfigure}
\begin{subfigure}{0.45\textwidth}
    \includegraphics[width=\textwidth]{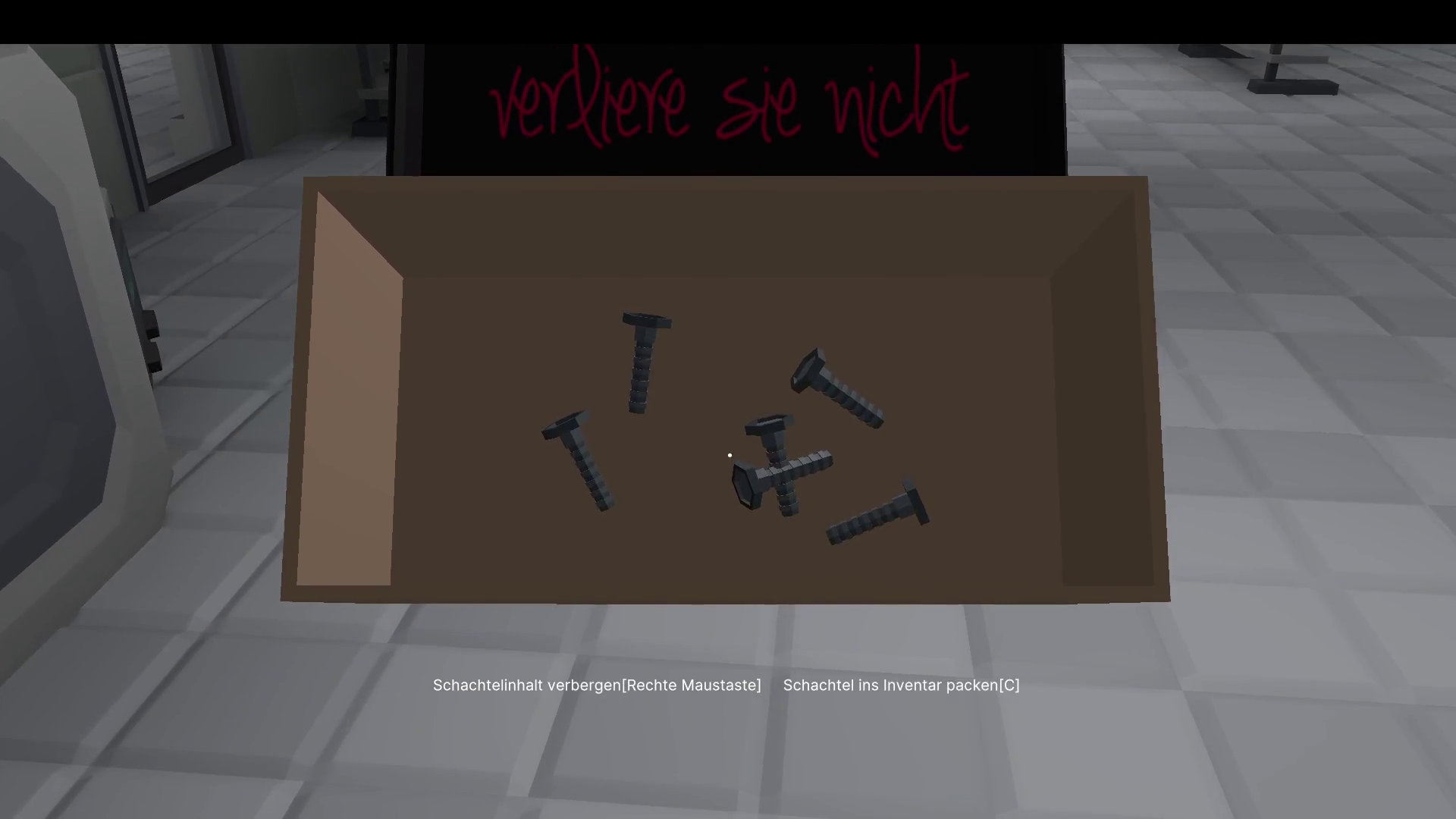}
    \Description[A box with six nails inside]{A screenshot of the game showing a box with six nails inside. It's representing the insistent questioning deceptive pattern}
    \caption{Nagging: Insistent Questioning}
    \label{fig:insistent-questioning}
\end{subfigure}
\begin{subfigure}{0.45\textwidth}
    \includegraphics[width=\textwidth]{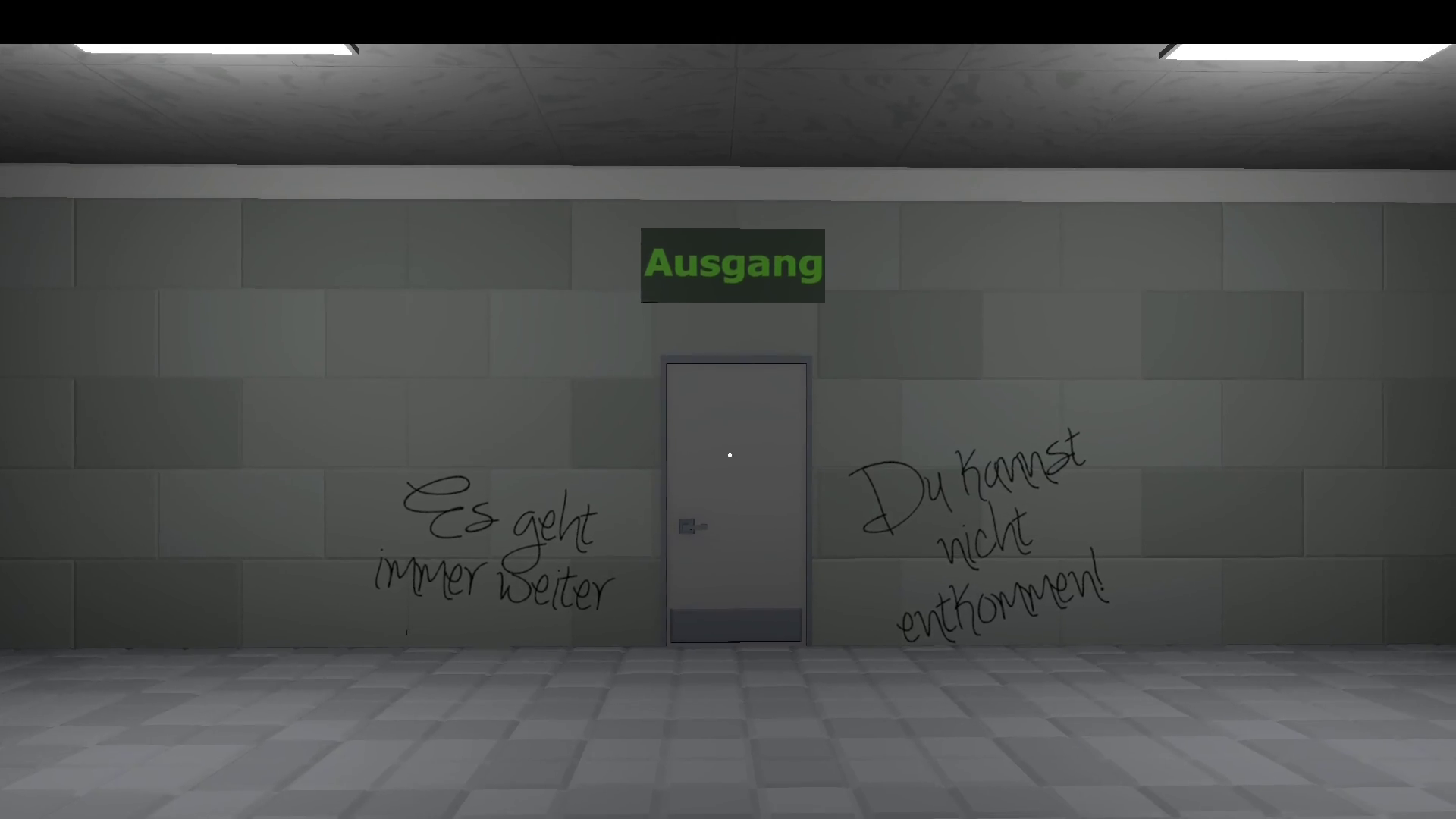}
    \Description[A room with a closed door, marked as exit and written messages on the wall]{A screenshot of the game showing a closed exit door. Markings on the wall indicate handwritten messages that it will always continue and that the player can't escape. It's representing the looping gameplay deceptive pattern}{}
    \caption{Forced Action: Looping Gameplay}
    \label{fig:looping-gameplay}
\end{subfigure}
        
\caption{Screenshots from the game representing each gamified deceptive pattern}
\label{fig:screenshots}
\end{figure}

\subsubsection*{\textbf{Preselection: Insensible Key Mapping}}
This deceptive pattern concept is defined as "any situation in which an option is selected by default prior to user interaction."~\cite{Gray.26.04.2018}. In our game, the narrator asks the player to accept a supposedly fine-tuned keyboard mapping to navigate and interact in the game, but the default keys are widely spaced and intentionally designed to inhibit ease of use. (\autoref{fig:insensible-key-mapping}). The player needs to realize that they should not just accept this default configuration but customize it. However, the interface is designed to make this customization cumbersome.

\subsubsection*{\textbf{Sneaking: Sneaky Shop}}
This deceptive pattern concept is defined as "attempting to hide, disguise, or delay the disclosure of information that is relevant to the user."~\cite{Gray.26.04.2018}. In our game, players must purchase work equipment from an AI-powered vending machine that sneaks items into the player's selection. If the player does not notice the sneak, the next step is a machine that checks if the player bought the correct items and discards them all if the sneak was successful (\autoref{fig:sneaky-shop}). The player cannot prevent the sneaking from happening, but if they are careful, they can undo it immediately and avoid having to redo the entire task.
    
\subsubsection*{\textbf{Hidden Information: Walls of Text}}
This deceptive pattern concept is defined as "options or actions relevant to the user but not made immediately or readily accessible."~\cite{Gray.26.04.2018}. In our game, players must read four boards containing lab rules with long texts to figure out a combination for setting eight levers placed around the room (\autoref{fig:walls-of-text}). While in the real world, examples like long lists of terms of use can often be skipped without reading, the game requires the player to read the entire text to find the correct configuration.

\subsubsection*{\textbf{Aesthetic Manipulation: Winding Hallway \& Shortcut}}
This deceptive pattern concept is defined as "any manipulation of the user interface that deals more directly with form than function."~\cite{Gray.26.04.2018}. In our game, the player enters a wide and long room with a narrow lighted path in the middle that seems to lead the user directly to the next room (\autoref{fig:winding-hallway}). Otherwise, the room is completely dark. The obvious choice for the player is to follow the lit path. However, the lighted path turns a corner just before the player thinks he has reached the next room, leading to a winding hallway with lots of advertisements on the walls and a much longer walk than expected. However, if the player explores the dark areas of the room, they may find a shortcut that the narrator tries to convince the player is not worth exploring. Ultimately, the player is free to take either the long or the short path in the game, both of which will allow them to continue, with the short path being darker and less appealing, but much faster.

\subsubsection*{\textbf{Obstruction: Obstacle Onslaught}}
This deceptive pattern concept is defined as "Making a process more difficult than it needs to be, with the intent of dissuading certain action(s)."~\cite{Gray.26.04.2018}. In our game, the player is confronted with a seemingly empty room with an immediately visible exit (\autoref{fig:obstacle-onslaught}). As the player attempts to get to the exit, a series of obstacles appear to prevent the player from advancing, combined with distractions designed to make the player think there might be a shorter path. The obstacles consist of a simple jump-and-run section, an invisible maze, a more difficult jump-and-run section, a series of unnecessary questions, and a search puzzle where the player must find a key. The distractions are teleporters that supposedly take the player to the goal. To overcome this deceptive pattern, the player must resist the distractions and survive each obstacle to reach the next room.

\subsubsection*{\textbf{Nagging: Insistent Questioning}}
This deceptive pattern concept is defined as "a redirection of expected functionality that persists beyond one or more interactions"~\cite{Gray.26.04.2018}. In our game, the player is initially given a set of objects, six small screws (\autoref{fig:insistent-questioning}). The narrator instructs the player to take care of them during the lab tour. But already in the next room, the narrator will try to convince the player to use the screws to repair things like squeaky doors or inefficient teleoperators. If the player trusts the narrator, he will end up with too few screws to get to the last room and will have to go back to get more. Therefore, to overcome the cheating pattern, the player must ignore the narrator's requests and guard the screws until the end of the game.

\subsubsection*{\textbf{Forced Action: Looping Gameplay}}
This deceptive pattern concept is defined as "Requiring the user to perform a certain action to access (or continue to access) certain functionality."~\cite{Gray.26.04.2018}. In our game, the narrator does not let the player leave at the end of the game, but instead sends them back to the beginning because the player "did not generate enough data" for the narrator's experiment (\autoref{fig:looping-gameplay}). While this may sound like a plausible explanation at first, it will happen again on repeat, forcing the player to replay a portion of the game over and over again. The only available countermeasure against Forced Action is not to use the service in the first place. Similarly, the only available remedy for repetition is to stop playing the game altogether.

%% file: 04_exploratory-study.tex
\section{Laboratory Gameplay Study (Study \#1)}
We conducted an exploratory laboratory gameplay study aiming to investigate how participants react to the gamified deceptive patterns. 
\subsection{Method}

Overall, the leading research questions were 

\begin{enumerate}
        \item \textbf{RQ1:} What are \emph{Influencing Factors}, which need to be considered for the design of such gamified deceptive patterns so that they can be effective to raise awareness?
        \item \textbf{RQ2:} What are the \emph{Motivations and Driving Forces} which might help to explain the behavior of users playing the game and falling or resisting a certain gamified deceptive pattern?
\end{enumerate}

\subsubsection{Apparatus}
We implemented the game \emph{Trickery} as a first-person game using \emph{Unity 2022.3.1f1}. The game consists of seven rooms, each representing a gamified deceptive pattern, in the order they appear in \autoref{sec:game-adapted-dark-patterns}. The game was designed to run on low-power devices to increase its potential applicability in various technical settings, such as privacy awareness training programs. We used low-poly assets from Synty \footnote{https://syntystore.com/, last accessed \today} to create the levels, as well as generative AI (\emph{elevenlabs.io}) for the voice of the narrator. Participant sessions were recorded in \emph{Zoom}\footnote{https://zoom.us/, last accessed \today} and transcribed using \emph{Adobe Premiere Pro}\footnote{https://www.adobe.com/products/premiere.html, last accessed \today}.

\subsubsection{Participants}
As an exploratory study, we selected a limited number of participants with the goal of conducting in-depth gameplay and interviews. In total, ten people (4 female, 5 male, 1 non-binary) participated in our study. Participants ranged in age from 22 to 32, with a median age of 26. Participants were selected through convenience sampling and advertisements for the study on social media platforms. The selection criteria ensured that participants had no prior knowledge of the game. The study was advertised and introduced as a video game study with no mention of deceptive patterns. We chose this approach to reduce pre-selection bias as well as demand characteristics. For example, participants may have been suspicious of the game from the start, knowing that the research topic was deceptive patterns. Of course, participants were fully debriefed after the study, see \ref{sec:ethics}. Participants were compensated with a 30€ gift card.

\subsubsection{Ethical Considerations} 
\label{sec:ethics}
Due to the lack of an ethics review board (ERB) at the host university at the time of planning, there was no direct way to involve such a ERB. Instead, the authors carefully followed the ethics checklists and guidelines of respected ethics boards. 
In particular, it was ensured that participants knew before they came to the study that they would be playing a video game, as playing such a game can induce a certain amount of stress and frustration. In addition, they were properly informed of the study procedures upon arrival and were fully debriefed at the end of the study. They were informed about the duration of the study (approximately 90-120 minutes), the procedure, and what data would be recorded during the session (game play, audio recordings of the game session, and interview). They were informed that they could withdraw from the study at any time.

\subsubsection{Procedure}
Participants were invited to our laboratory, where they were first informed about the study and the information to be recorded (see \autoref{sec:ethics} for ethical considerations). They were asked to give their consent to the data collection and study procedures. Participants were then presented with the game and asked to begin playing while thinking aloud about their actions, experiences, and feelings. During the game session, the two study directors only intervened if participants got stuck for a long time or if an error occurred in the game. After completing the game, participants were debriefed and informed about the context of the game, i.e., that the purpose of the game was for the narrator to keep the player in the game as long as possible and that similar manipulation techniques are common on websites. A semi-structured interview was then conducted. During this interview, the interviewer asked the participant to recap each room one at a time, with each room reflecting a gamified deceptive pattern. For each room, participants were asked about (1) their overall state of mind, (2) the behavior they thought was expected of them, (3) how they reacted to unexpected challenges (i.e., if completing the task was not as easy as it may have seemed). We then explained and revealed the true nature of the room/game deceptive pattern and asked participants if they would have changed their behavior given this knowledge. Finally, we asked participants if such a situation reminded them of a real-world scenario when using digital applications such as websites or apps. After this retrospective walkthrough of all seven gamified deceptive patterns, we asked participants how their perspective of the narrator changed over the course of the game, if they learned anything, and what they liked or disliked about the game.
The gameplay portion lasted an average of one hour. The interview portion lasted between 45-60 minutes.

\subsubsection{Analysis}
The recordings of both the gameplay sessions and the interviews were transcribed after the study was completed. To analyze the data qualitatively and to identify patterns in the participants' statements, we performed a thematic analysis, following the steps described by \citeauthor{braun2006using}~\cite{braun2006using}. First, the transcripts were read thoroughly to gain a comprehensive understanding of them. Bullet points and possible codes that summarized the features contained in the data were noted. In the second step, two researchers independently examined and coded the data. The two researchers then worked together to generate themes from the codes, iterating and adjusting the themes until they were internally homogeneous. In a final iteration, the themes were discussed and refined with a third researcher. This resulted in two main themes with multiple sub-themes. These themes are described in \autoref{sec:results}.

\subsection{Results: Laboratory Gameplay Study}
\label{sec:results}
This section presents the results of our analysis. Since both the game and the interview were conducted in a language other than English, we translated the participants' statements into English for the purposes of reporting. Participant statements are presented in quotation marks, with each quoted participant identified by their participant ID.

    \subsubsection{Overall Gameplay Experience}
 Overall, all participants were able to complete the game. The gameplay loop in the final room was endured 4.5 times (median) before participants decided or understood that only exiting the game would stop the loop. All participants showed some amount of trust in the narrator as they initially accepted the suggested "fine-tuned keyboard mapping". The participants' behavior over the course of the game showed this trust to decrease over time, but even at the very end of the game, participants followed the instructions to at least some extent. This was particularly evident in the obstruction room, where participants repeatedly tried to use the teleporters even though they were clearly part of the narrator's obstruction. 
    Since the players expected to simply play an entertaining video game, most of them were visibly frustrated with the game and also stated during the interview that they did not enjoy playing the game. Interestingly, this was reversed after being informed about the purpose of the game, which seemed to make the frustration and challenges worthwhile. Most importantly, the game made participants aware of how easily they can be manipulated when led by an antagonistic narrator. Accordingly, not to trust blindly and to think carefully about decisions was mentioned, e.g. by P1: "Maybe I should not always act so hastily, even with all these decisions. And just because someone tells me to do this or that", as well as P10 "Yes, maybe that sometimes you should listen to your intuition and even if you're put in a situation and think I'm doing this for other people, that you should stop when it's no longer good for you."
    
    \subsubsection{Player Motivations and Driving Forces}
    Our thematic analysis revealed several motivating factors and driving forces that participants used to justify their behavior. These factors may not only be used to improve upon gamified deceptive pattern design but also provide a different outlook on how users behave when faced with deceptive patterns in general.
    \begin{itemize}
        \item[\textbf{DF 1: }]\emph{Curiosity}
   Curiosity appears at various points in the game, motivating players to do things that sometimes hinder their progress by falling for the gamified deceptive patterns. For example, three participants justified their decision to accept the Insensible Key Mapping with curiosity, stating that they "want to know what this preferred key mapping is" (P01). Similarly, several participants used curiosity about new clues to justify their decision not to quit the game during the looping gameplay, even though the clues on the wall indicated that they should. For example, P04 stated, "I could keep playing until no new text was added. We found that visible changes, such as the added hints on the wall or the color change of the teleporters (one of the strange behaviors of the teleporters), increased participants' curiosity and distracted them from the countermeasures to the deceptive patterns available to them. 
        
        \item[\textbf{DF 2: }]\emph{Frustration and Resilience:}
  At several points in the game, participants' behavior was influenced by frustration. For example, participants were frustrated by the Insensible Key Mapping, which motivated them to open the menu and change the key mapping. Several participants expressed frustration during Walls of Text and decided to switch between different texts or skip text passages in the hope of finding the solution without reading the entire text. In the later game, Obstacle Onslaught, several participants were so frustrated by the narrator's persistent manipulation that they deliberately ignored the teleporters he was trying to get them to use. In response to the narrator's request for a screw to fix a teleport, P01 said, "No, not anymore!" (as in they won't give him screws anymore), followed by "I probably wouldn't get anywhere either. Interestingly, things that frustrated some participants did not seem to bother other, more resilient participants. Two participants managed to complete the game without changing the key mapping to a more common one, one participant approached Walls of Text by simply reading the texts, and several participants exhausted all available clues on the wall in Looping Gameplay despite knowing they could quit the game.  
                
        \item[\textbf{DF 3: }]\emph{Cost-Benefit Consideration:}
  Another driving force behind participants' decisions was cost-benefit considerations. In particular, when asked why they did or did not give away screws in Insistent Questioning, participants expressed that they decided on a question-by-question basis whether the benefit of giving away the screw was great enough. This was even evident in some participants' out loud thinking during the game, such as P09's "Well, if it makes it easier for me, then sure!" Another segment of the game where cost-benefit considerations played a large role was Obstacle Onslaught, where several participants started using the teleporters when they found the obstacles too difficult. P06 said, "I didn't go to the teleport right away, I looked at what was in front of me. Then I tried it if the path was not solvable."
                
        \item[\textbf{DF 4: }]\emph{Conformity with Player Expectations:}
                We found that some in-game decisions were influenced by how participants expected video games to behave. For example, during the Insensible Key Mapping, participants expressed that they expected the narrator to help them and initially trusted him (P02: "I said I trust the game. Yeah, it's going to know what's good for you.") This also applied to the key mapping itself, with players expecting the key mapping to conform to first-person game conventions (P06: "At first, you would approach this benevolently and think 'the narrator only means well,' it's going to be the right key mapping. And that standard is WASD.")
                
                A similar expectation expressed by the participants is that the game has a solvable ending. Many participants were surprised by the last required action, i.e. leaving the game, with only two participants expressing that they had seen something similar in games before. The other participants behaved in a way that showed that they expected an in-game solution. P02 summed up this expectation very well: "You see this 'Exit' sign and you think, 'Yes! Thank you! I can get my carrot here. My dopamine. The game is over. Pat yourself on the back.'" When describing how he felt about the lack of an in-game ending, P02 said, "I'm unhappy. Definitely. I was expecting one thing and I did not get it.
                
                The most prominent example of players' expectations shaping their behavior, however, came from Winding Hallway \& Shortcut. Namely, the dark parts of the room and the dark hallway representing the shortcut reminded participants of horror games. When asked about the lamp breaking to indicate to participants that there was more to the room than the lit path, P04 said, "I'm not a fan of horror games. That's why I don't turn around for things like that. Similarly, P07 said, "This looks a bit scary. You think, 'Okay, this is what a classic horror game might look like.'" This expectation that there might be horror elements in the game was so strong that even participants who were told by the researchers that there were no horror elements in the game did not go to investigate the dark areas of the room.
    \end{itemize}

    \subsubsection{Influencing Factors for Adapting Deceptive Patterns into Gameplay}
    In our analysis, we discovered three factors that influenced how the participants experienced the gamified deceptive patterns. Keeping these influencing factors in mind may help researchers understand which parts of their implementation successfully communicate the intended message, and which parts may get lost to the players.

    \begin{itemize}
        
        \item[\textbf{IF 1: }]\emph{Mapping fidelity between gamified deceptive pattern and real-world deceptive pattern example:}
 When asked which deceptive pattern examples participants associated with the gamified version, we found that when the latter used graphical user interface (GUI) elements to convey the deceptive pattern concept (e.g., Insensible Key Mapping, Sneaky Shop, or Walls of Text), participants were able to refer to real-world deceptive pattern examples and match the corresponding deceptive pattern concept (e.g., Preselection, Sneaking, and Hidden Information, respectively). Those gamified deceptive patterns that were not implemented through GUI elements, but were adapted by manipulating the actual game world (e.g., Winding Hallway \& Shortcut, Obstacle Onslaught, or Looping Gameplay) yielded less clear associations. While this suggests that presenting deceptive patterns in a familiar form may lead to better in-game recognition, we cannot conclude that this led to greater awareness of or resistance to the deceptive pattern concepts.  
        
       \item[\textbf{IF 2: }]\emph{Ambiguity of deceptive pattern concepts and real-world deceptive pattern examples:}
 Even while researching possible deceptive pattern concepts to incorporate into gamified deceptive patterns, we noticed that existing deceptive pattern taxonomies sometimes struggle with the ambiguity between different deceptive pattern concepts (e.g., Sneaking and Hidden Information). This is especially the case when trying to design a gamified deceptive pattern that should represent only one specific concept. In our game, an unintentional aesthetic manipulation was observed in the Insensible Key Mapping level. The narrator's positive wording ("a customized control") creates a false hierarchy between using the preselected control and denying it, effectively using multiple deceptive pattern concepts at once. Several participants used this phrase to justify accepting the pre-selected controls. Similarly, for some participants, the focus of looping gameplay was not on the forced action, i.e., being forced to do something you did not want to do (repeat the game) in order to achieve your goal (finish the game), but on the repetition itself, which would place the gamified deceptive pattern concept closer to nagging.
        
        \item[\textbf{IF 3: }]\emph{Balance between gamified deceptive patterns and enriching game mechanics:}
     During the analysis, we noticed that game mechanics introduced to make the game more enjoyable, or to enrich the game, seemed to be in mental conflict with the deceptive patterns built into the game. For example, in our implementation of Walls of Text, several participants felt exhausted or annoyed by at least one puzzle they had to solve to get the correct lever combination. Similarly, in Obstacle Onslaught, we introduced two jump-and-run segments that some participants struggled to complete due to the difficulty of controlling the jumps, while others responded positively, stating that "at least this has finally become a video game" (P05). In both Walls of Text and Obstacle Onslaught, participants focused more on the enriching elements of the game when describing their experience, overshadowing the gamified patterns of deception (e.g., having to search for hidden information or face obstacles). In addition, we found that repeating a gamified deception pattern to enrich the experience too often (e.g., showing four walls of text instead of two, using four obstacles in Obstacle Onslaught instead of three) can be detrimental to the learning experience. While frustration is part of the game design, repeating the required countermeasure so often can distract participants from the actual gamified deceptive pattern.
    \end{itemize}
    
\subsubsection{Relation to Real-World Deceptive Patterns}
In terms of participants' ability to relate the gamified deceptive patterns to real-world experiences in digital applications, results varied. For the \emph{Preselection} pattern, six participants compared it to a cookie banner, interestingly describing it as something that distracts and annoys the user by preventing them from actually visiting the website (here: from easily navigating the game). Regarding \emph{Sneaking}, five participants related this to very similar experiences with web stores or program installation procedures that tried to sneak in other items. For example, P8 said, "Amazon does that, for example. That if you don't have Prime, it automatically says get Prime, and then it's also pre-selected that you can buy a Prime subscription with it. \emph{Hidden Information} was very clearly associated with \emph{Terms and Conditions} of websites or applications (nine out of ten respondents). Interestingly, \emph{Aesthetic Manipulation} was most often (seven participants) compared to annoying ads on websites. Here, the actual manipulation was not considered by participants when looking for real-world comparisons. Instead, the use of ads in the long hall to annoy the user seems to have overshadowed this game element. While \emph{Obstruction} was clearly the most challenging and potentially annoying room in the game, it made it difficult for participants to find direct comparisons. This raises the question of whether \emph{Obstruction} may be one of those deceptive patterns that is not well recognized or simply less common in real-world environments. The results for \emph{Forced Action} are very similar, as the concept is also quite comparable in many situations. The \emph{Nagging} deception pattern, which was placed along the entire game experience, was interestingly related to in-game currencies or in-app purchase options.  Nevertheless, given the explanation of the pattern, three participants were able to compare it to real-world scenarios, e.g., P9 stated, "For example, you grant access to the gallery or to contacts and basically you also grant a lot of rights [without much thought].

Overall, participants' interpretations of the deceptive patterns were mixed and ambiguous. Due to the design of the study, we could not rule out the possibility that playing the game and being introduced to the idea of the gamified deceptive pattern may have introduced a certain bias in participants to basically try to find 1:1 examples in the real world. To explore this transfer potential in more detail, we conducted a second study, an online survey, that focused specifically on the potential transfer learning capabilities of the gamified deceptive patterns.

%% file: 04_survey-study.tex
\section{Online Survey Study (Study \#2)}
\label{sec:online_study}
Consequently, in this online survey, we focused on the following research question:

\begin{quote}
    \textbf{RQ3: } Do users actually find our gamified deceptive patterns helpful to understand the real deceptive pattern concepts from the literature?
\end{quote}



\subsection{Method}
In order to reach a larger audience we opted for an online survey. Here, we presented participants with a short video clip featuring \emph{Trickery} gameplay for each gamified deceptive pattern (ca. 1 minute each) and a textual explanation of that gameplay. Following each video, we presented participants with a definition of the corresponding real-world deceptive pattern concept adapted from \cite{Gray.26.04.2018}. They then had to rate each gamified deceptive pattern on a 5-point semantic differential scale with anchors 1 (\emph{helpful to understand the deceptive pattern concept}) and 5 (\emph{hindering to understand the deceptive pattern concept}).


\subsubsection{Survey Procedure}
The survey was created on a self-hosted LimeSurvey platform \footnote{\url{https://www.limesurvey.org/}} and published via SurveyCircle \footnote{\url{https://surveycircle.com/de}}. It was conducted in German language. On the initial page of the survey, we explained the topic of our survey, defined the term deceptive patterns, provided the rough narrative of the \emph{Trickery} game, and an overview of how the rest of the survey would proceed. We informed participants about the duration of the survey (approx. 15-20 minutes). Participants were informed that their survey responses would be a) collected, stored, and analyzed anonymously, b) that their responses would be used for analysis and presentation of results, and c) that they could quit participation at any time. After providing consent for data collection and presentation, participants were presented with the first (out of seven) gamified deceptive patterns and corresponding deceptive pattern definition. The order of the deceptive patterns was randomized between participants. For each gamified deceptive pattern, a short video of roughly one-minute length was produced which explained the gameplay events concisely. It was ensured that the video commentary was purely descriptive and aimed to convey what was happening in the game, rather than why it was happening (i.e., we refrained from explaining the narrator's intent behind the laboratory setup). This way, watching the videos would most closely match a playing experience without a prior explanation of the precise motivation behind deceptive patterns (i.e., prior to a debriefing in an educational setting). Following each video, we provided a definition of the corresponding deceptive pattern concept adapted from \cite{Gray.26.04.2018}. We slightly altered the exact wording for each definition to directly address the participants and evoke a more personal experience. The participants were then asked to rate the helpfulness on the 5-point semantic differential scale. Specifically, they had to complete the following sentence through their rating
\begin{quote}
\emph{The presented video sequence was [1-helpful | 5-hindering] for my understanding of the deceptive pattern [title of deceptive pattern concept]}.     
\end{quote}
Participants were then asked to provide up to three deceptive pattern examples. 

The survey concluded with questions on demographic data, specifically regarding the participants' age, their gender identity, and how often the participants play video games on a scale of \emph{Daily}, \emph{Multiple Times a Week}, \emph{Once a Week}, \emph{Once a Month}, \emph{Fewer}, \emph{Not at all}.

\subsubsection{Participants}
34 participants (16 female, 17 male, 0 non-binary, 1 preferred not to say) completed the survey. Participants' ages ranged from 18 to 56 years (median=30.24y, sd=9.04). 23 participants (67.6 percent) reported to play video games at least once a week. We recruited participants via convenience sampling and use of the platform \emph{SurveyCircle} \footnote{\url{https://surveycircle.com/de}}, including spreading the survey via SurveyCircle's social media presences and social media groups. Participants could enter a raffle for one of ten 15€ gift cards upon completion of the survey and were informed about that possibility at the beginning.

\subsubsection{Analysis} 
Our analysis included two steps: 
\begin{enumerate}
    \item Evaluate the provided deceptive pattern examples to determine whether the participants understood the deceptive pattern concept.
    \item Analyze the perceived helpfulness of the gamified deceptive pattern of those users who had provided at least one fitting deceptive pattern example (see step 1).
\end{enumerate}

The reasoning behind this approach was as follows. As we provided a textual definition of each deceptive pattern concept to avoid biases and without the ability for participants to ask questions, we assumed that not all participants would understand the definition correctly. However, if we had included exact examples of deceptive patterns (which certainly would have helped participants to understand the idea), we would have introduced a bias for the helpfulness rating on our semantic differential scale: Participants would most likely then compare the gamified deceptive pattern to any provided example by us.

Therefore, we asked participants, after having done the helpfulness rating, for specific examples of their own, which we could then use to validate their level of understanding.


Two researchers independently read all provided examples and decided whether or not a participant was able to provide a matching example for a given deceptive pattern. Out of 34 participants, between 15 and 22 participants remained in the sample of the respective deceptive pattern due to being able to provide at least one example which could be aligned to the deceptive pattern concept from \cite{Gray.26.04.2018}.


We then analyzed the remaining participants rating scale data (helpfulness) with a one-sample Wilcoxon signed-rank test. The test checks whether the median of a distribution, such as the responses to a semantic differential, is significantly different compared to a predefined value. We compared our scores to the neutral value of 3 (neither helpful nor hindering) on the 5-point semantic differential. 

\subsection{Results: Survey Study}
\begin{figure}[h]
    \includegraphics[width=\linewidth]{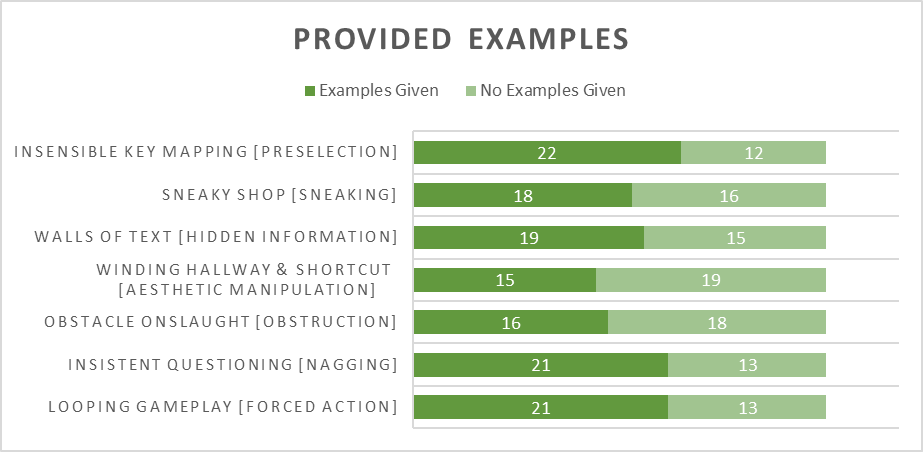}
    \Description[A graph showing the distribution of participants who could provide an example vs. those who could not for each deceptive pattern concept]{A graph showing how many users provided examples for which deceptive pattern concept. It shows that 22 participants could provide examples for insensible key mapping / preselection, while 12 could not. 18 provided examples for sneaky shop /sneaking, while 16 could not. 19 provided examples for Walls of text / hidden information, while 15 could not. 15 provided examples for winding hallway \& shortcut / aesthetic manipulation, while 19 could not. 16 provided examples for obstacle onslaught / obstruction, while 18 could not. 21 provided examples for insistent questioning / nagging, while 13 could not. 21 provided examples for looping gameplay / forced action, while 13 could not.}
    \caption{Distribution of participants with and without matching examples provided for each gamified deceptive pattern}
    \label{fig:provided_examples}
\end{figure}

As briefly mentioned, 15 to 22 (out of 34) participants were able to provide matching deceptive pattern examples. The precise values are provided and visualized in \autoref{fig:provided_examples}. 

\begin{figure}[h]
    \centering
    \includegraphics[width=\linewidth]{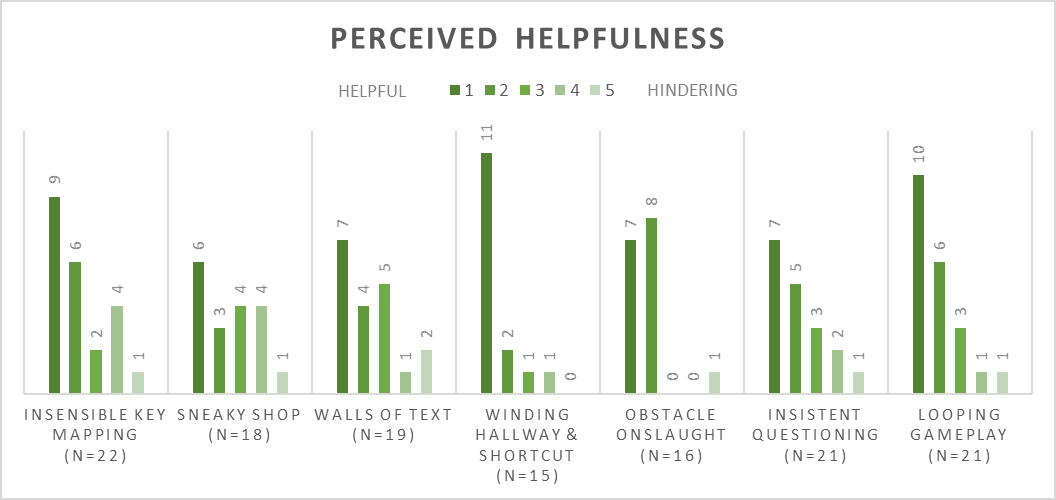}
    \Description[Histogram of the distribution regarding the results of the semantic differential scale]{Histogram showing that for insensible key mapping, 9 participants rated it with 1, i.e. helpful, 6 with a 2, 2 with a 3, 4 with a 4 and 1 with a 5. For Sneaky shop, 6 participants rated it with 1, i.e. helpful, 3 with a 2, 4 with a 3, 4 with a 4 and 1 with a 5. For Walls of Text, 7 participants rated it with 1, i.e. helpful, 4 with a 2, 5 with a 3, 1 with a 4 and 2 with a 5. For Winding Hallway, 11 participants rated it with 1, i.e. helpful, 2 with a 2, 1 with a 3, 1 with a 4 and 0 with a 5. For Obstacle Onslaught, 7 participants rated it with 1, i.e. helpful, 8 with a 2, 0 with a 3, 0 with a 4 and 1 with a 5. For Insistent Questioning, 7 participants rated it with 1, i.e. helpful, 5 with a 2, 3 with a 3, 2 with a 4 and 1 with a 5. For Looping Gameplay, 10 participants rated it with 1, i.e. helpful, 6 with a 2, 3 with a 3, 1 with a 4 and 1 with a 5.}
    \caption{Histogram of perceived helpfulness of each gamified deceptive pattern}
    \label{fig:perceived_helpfulness}
\end{figure}

The results of the Wilcoxon signed-rank tests revealed that the perceived helpfulness deviates significantly from the neutral median (3) for \emph{Insensible Key Mapping}, \emph{Winding Hallway \& Shortcut}, \emph{Obstacle Onslaught}, \emph{Insistent Questioning}, and \emph{Looping Gameplay}. The Wilcoxon signed-rank test did not reveal a significant deviation from the neutral median for \emph{Sneaky Shop} and \emph{Walls of Text}. Descriptive statistics, as well as the test statistics for the Wilcoxon signed-rank tests for each gamified deceptive pattern, are presented in \autoref{tab:perceived-helpfulness}. The distribution of responses for perceived helpfulness is shown in \autoref{fig:perceived_helpfulness}. We can see that besides \emph{Sneaky Shop} and \emph{Walls of Text} the remaining five deceptive patterns were rated as helpful, with the difference compared to the neutral value being significant.

\begin{table}[h]
\resizebox{\linewidth}{!}{%
\begin{tabular}{lccccccc}
                                & \textbf{N} & \textbf{Mean} & \textbf{Standard Deviation} & \textbf{Median} & \textbf{IQR} & \textbf{Z} & \textbf{p-Value}  \\
\textbf{Insensible Key Mapping [Preselection]}           & 22         & 2.18          & 1.296                       & 2.0             & 2            & -2.596     & 0.009*            \\
\textbf{Sneaky Shop [Sneaking]}               & 18         & 2.50          & 1.339                       & 2.5             & 3            & -1.647     & 0.100             \\
\textbf{Walls of Text [Hidden Information]}     & 19         & 2.32          & 1.336                       & 2.0             & 2            & -1.919     & 0.055             \\
\textbf{Winding Hallway \& Shortcut [Aesthetic Manipulation]} & 15         & 1.47          & 0.915                       & 1.0             & 1            & -3.361     & \textless{}0.001* \\
\textbf{Obstacle Onslaught [Obstruction]}            & 16         & 1.75          & 1.000                       & 2.0             & 1            & -2.954     & 0.003*            \\
\textbf{Insistent Questioning [Nagging]}                & 18         & 2.17          & 1.249                       & 2.0             & 2            & -2.368     & 0.018*            \\
\textbf{Looping Gameplay [Forced Action]}          & 21         & 1.90          & 1.136                       & 2.0             & 2            & -3.086     & 0.002*           
\end{tabular}%
}
\vspace{6pt}
\Description[Descriptive data of the semantic differential scale as well as the statistical test results]{Descriptive data of the semantic differential scale as well as the statistical test results}
\caption{Descriptive data and one-sample Wilcoxon signed-rank test results on the perceived helpfulness of a gamified deceptive pattern for a deceptive pattern concept. (N: Number of participants with matching examples, IQR: Interquartile Range, Z: Wilcoxon signed-rank test statistic). Significant results are indicated by an *. Mean / Median values on a scale of 1-5, 1=\emph{helpful}, 5=\emph{hindering}}
\label{tab:perceived-helpfulness}
\end{table}

%% file: 05_discussion.tex
\section{Discussion}
\label{sec:Discussion}
The results of the laboratory gameplay study and online survey study provide us with important cues into the design and implementation of gamified deceptive patterns. While these were designed to represent real-world deceptive pattern concepts and align with specific deceptive pattern examples, the results of our laboratory study reveal that several influencing factors play a role in how the the gamified versions are interpreted and which factors influence player behavior. Our online survey study provides insights into the perceived helpfulness of the gamified deceptive patterns to understand the overarching concept behind them.

First, we want to address the research question posed in our online survey study (\autoref{sec:online_study}) Do users actually find our gamified deceptive patterns helpful in understanding real deceptive pattern concepts from the literature (RQ3)?

Examining the results, we see that for five of the seven gamified deceptive patterns, the median is significantly different from the neutral value in the direction of \emph{helpful}. This indicates that participants who understood the underlying deceptive pattern concepts for \emph{Insensible Key Mapping}, \emph{Winding Hallway \& Shortcut}, \emph{Obstacle Onslaught}, \emph{Insistent Questioning}, and \emph{Looping Gameplay} would rate them as helpful. We infer that the conceptual design of these deceptive patterns is therefore consistent with the overarching concepts of deceptive patterns, implying that they are promising solutions for raising awareness --- tackling one of three countermeasures proposed by \citeauthor{BongardBlanchy.06282021}~\cite{BongardBlanchy.06282021}. 

Interestingly, \emph{Sneaky Shop} and \emph{Walls of Text}, two gamified deceptive patterns for which participants in the laboratory gameplay study were mostly able to provide matching examples, were not found to be particularly helpful. Comparing these results with the results of the laboratory gameplay study, specifically \emph{IF 1: Mapping fidelity between gamified deceptive pattern and real-world deceptive pattern example}, we see that the very patterns that reminded the gameplay study participants of specific real-world deceptive pattern examples due to their GUI-based representation were not perceived as helpful by the online survey study participants. This suggests that while a GUI-based representation may more closely resemble familiar deceptive pattern examples, it may not be the most helpful way to convey the overarching deceptive pattern concept, which may extend beyond such specific examples, underlining the importance of narrative elements~\cite{naul2020story} and intrinsic integration of learning content into the game world~\cite{habgoodMotivatingChildrenLearn2011a}. This finding is corroborated by the fact that although less than half of the participants in the online survey study were able to provide appropriate examples for \emph{Winding Hallway \& Shortcut} and \emph{Obstacle Onslaught} --- both of which represent their corresponding deceptive pattern concept by manipulating the game world rather than the GUI --- their perceived helpfulness is rated quite high.

Examining the Laboratory Gameplay Study Influencing Factor (RQ1) \emph{IF 3: Balance between gamified deceptive patterns and enriching game mechanics}, it is clear that serious game designers need to consider enriching game mechanics in their implementation of deceptive pattern analogies. A balance must be struck between engaging players through variety and novelty in the games, and raising awareness and increasing resistance. We hypothesize that the enriching game mechanics should be the easier parts of such a serious game, and the gamified deceptive patterns, which by design frustrate players, take on the more difficult parts - as evidenced by the driving force (RQ2) \emph{DF 2: Frustration and Resilience}. How this balance affects the success of gamified deceptive patterns in raising awareness is an exciting question for further research.

As a serious game concept, it is important to consider how the game would be presented to potential learners~\cite{kronhardtStartPlayingSerious2024}. In the case of our laboratory gameplay study, we did not reveal the theme of the game until the end of the gameplay session. This was done deliberately to prevent participants from changing their behavior based on this knowledge. Looking at the results, we believe that this is a reasonable and possibly more successful approach than revealing the game's theme before the gameplay session. As discussed in \emph{DF 4: Matching Player Expectations}, players have an initial trust in the game, and their behavior is strongly driven by \emph{DF 1: Curiosity} and \emph{DF 2: Frustration}. Revealing that the game is trying to manipulate them would foster distrust from the start, and thus could affect learning. Observant players might be able to figure out the necessary countermeasures without first experiencing the negative consequences of falling for the deceptive pattern, thus inhibiting the building of awareness. Therefore, future research should focus not only on the design of gamified deceptive patterns, but also on their integration into an educational environment where the concept can be used to its full potential. Alternatively, intrinsically integrating a debriefing element within the game which explains the connection between gamified deceptive patterns and their deceptive pattern concepts could prove useful~\cite{naul2020story, habgoodMotivatingChildrenLearn2011a}. For example, when trying to quit the game, the narrator could break the fourth wall and guide the player through the laboratory one final time, explaining the motivation behind each room. 

Throughout gameplay, participants gradually developed a relationship with the narrator, initially trusting their guidance but later realizing the narrator’s actions were increasingly manipulative. This shifting perception allowed players to experience the consequences of being deceived, possibly enhancing their understanding of deceptive patterns. While already proven useful as a narrative device in our studies, this dynamic between narrator and player could in itself be used in future work to educate about more sublte linguistic deceptive patterns, such as misleading voice prompts and restricted options~\cite{owensExploringDeceptiveDesign2022}, which could be applied in conversational user interfaces~\cite{mildnerListeningVoicesDescribing2024}.

\section{Limitations}
\label{sec:limitations}
Our work has several notable limitations. As discussed above, we observed instances where certain game mechanics overlapped with gamified deceptive patterns, as well as instances where the implementation of a deceptive pattern used deceptive pattern concepts that did not match the intended underlying deceptive pattern concept. While this overlap posed a challenge in isolating the specific effects in these particular cases, it allows us to explore the intricate relationship between game mechanics and gamified deceptive patterns. We addressed this limitation by designing the online survey study in a highly descriptive manner, decoupling the conceptual idea of our gamified deceptive patterns from their precise implementation by presenting them in video form.

As is often the case with exploratory studies, the results of our studies are context-dependent and apply primarily to the specific environment and user group studied (e.g., consisting mostly of people familiar with video games), especially given the small sample size in our study. Consequently, the generalization of our results should be approached with caution. Nevertheless, our work presents intriguing hypotheses that provide guidance for future research, design, and implementation of gamified deceptive patterns. The designs presented and the narrator-based gameplay concept also provide a framework for researchers to more quantitatively evaluate the success of this concept in raising awareness of and increasing resistance to deceptive patterns. Despite the small sample size of the online survey study, our quantitative analysis yielded statistically significant results indicating that most of the gamified deceptive patterns were found to be helpful. 

Additionally, the presence of certain usability issues within the game during the laboratory gameplay study may have inadvertently increased participants' frustration levels. It is important to note, however, that although participants expressed frustration during gameplay, after receiving a comprehensive debriefing explaining the context of the study, they appeared to attribute these moments of frustration to potentially deceptive pattern manipulations, thus limiting the impact of the usability issues on participants' opinions regarding our research questions. 
 From a methodological perspective, there are a few additional issues worth mentioning here. The laboratory gameplay study asked participants to think aloud, which is common in usability testing but less common in experimental setups. Here we found that it resulted in a very natural "live game commentary" that did not seem to cause unnecessary cognitive load. We were also careful not to encourage participants to think aloud if they had been silent for some time. For the gameplay study, we did not explicitly ask participants about their gaming experience or whether they were familiar with any of the closely related games, such as The Stanley Parable. However, none of the participants mentioned any of these games during the session. Also, we have to acknowledge that the experience for the participants in the survey study is quite different from playing the game. Nevertheless, we believe that for the particular focus here, this was the best choice in order to reach a larger number of participants and not have certain aspects of the game not related to the direct transfer of the gamified deceptive patterns influence the participants.

The gameplay design and studies concluded before the release of the most current, comprehensive ontology by \citeauthor{gray_ontology_2024}~\cite{gray_ontology_2024}. However, the ontology encompasses and integrates the taxonomy used in \textit{Trickery}~\cite{Gray.26.04.2018} with other taxonomies from research and regulatory bodies~\cite{gray_ontology_2024}. Therefore, a direct mapping between our gamified deceptive patterns and the new ontology is possible, albeit partially at different levels. Briefly speaking: \textit{Obstruction}, \textit{Sneaking}, \textit{Interface Interference}, and \textit{Forced Action} remain high-level patterns in the new ontology. Similarly, \textit{Hidden Information}, \textit{Aesthetic Manipulation}, and \textit{Preselection} remain subtypes of \textit{Interface Interference} as meso-level patterns, partially under different names: \textit{Hidden Information}, \textit{Manipulating Visual Choice Architecture}, and \textit{Bad Defaults}, respectively. The remaining deceptive pattern concept from \textit{Trickery}, \textit{Nagging}, is mapped as a meso-level pattern under \textit{Forced Action} in the new ontology~\cite{gray_ontology_2024}. Additionally, the newly introduced high-level pattern \textit{Social Engineering} is missing from the current selection of gamified deceptive patterns in \textit{Trickery}. While this mapping allows us to situate our work within the most current research, future work could expand our gamified deceptive patterns to include \textit{Social Engineering} and implement further meso-level patterns for the other high-level patterns only represented as one gamified deceptive pattern in our game. This would also increase the replayability of the game, as different gamified meso-level patterns for certain high-level patterns could randomly be employed in each playthrough.

%% file: 06_conclusion.tex
\section{Conclusion}
In this paper, we presented the narrator-based serious game \emph{Trickery} and its implementation along with seven gamified deceptive patterns designed to raise awareness of deceptive patterns and ultimately help strengthen resistance to them through the consequences of player actions.
In an initial exploratory study, we identified key influencing factors (RQ1) for incorporating deceptive pattern concepts into gameplay, as well as player motivations and driving forces (RQ2) that influence player behavior. In a subsequent online survey study, we also found evidence that most gamified deceptive patterns are effective for understanding and relating to the overarching deceptive pattern concepts (RQ3). However, two analogies that used direct mapping through a GUI-based approach, similar to existing deceptive pattern examples, were found to be less helpful.

Based on these results, we propose three new hypotheses for future research on gamified deceptive patterns. Based on our work, future research could: 1.) Explore the impact of the fidelity of gamified deceptive patterns on deceptive pattern awareness in more detail, particularly building on the results of our online survey. 2.) explore the balance and interplay between the fun provided by enriching game mechanics and the frustration inherent in gamified deceptive patterns, and 3.) explore how gamified deceptive patterns can be integrated into instructional processes while maintaining the important driving forces of player expectation, curiosity, and frustration.